# Abundance Determinations from Gamma Ray Spectroscopy

Reuven Ramaty

*Laboratory for High Energy Astrophysics, Goddard Space Flight Center, Greenbelt, MD 20771*

**Abstract.** Gamma ray emission lines resulting from accelerated particle bombardment of ambient gas can serve as an important spectroscopic tool for abundance determinations. The method is illustrated by considering the gamma ray line emission observed from solar flares. The observation of similar gamma ray lines from Orion suggests the existence of large fluxes of low energy Galactic cosmic rays. The role of these cosmic rays in the nucleosynthesis of the light isotopes is discussed.

## 1. Introduction

The interactions of accelerated particles with ambient matter produce a variety of gamma ray lines following deexcitations in both the nuclei of the ambient medium and the accelerated particles. Astrophysical deexcitation line emission produced by accelerated particle interactions has so far been observed from solar flares (e.g. Chupp 1990; Share & Murphy 1995) and the Orion molecular cloud complex (Bloemen et al. 1994). For a general review of astrophysical gamma ray line emission see Ramaty & Lingenfelter (1995). The solar gamma ray line observations have many applications, including the determination of solar atmospheric abundances (Murphy et al. 1991; Ramaty et al. 1995a; Ramaty, Mandzhavidze, & Kozlovsky 1996a). The Orion observations, even though much less detailed, have nonetheless revealed the existence of large fluxes of low energy cosmic rays in this nearest region of recent star formation (e.g. Ramaty 1996). If such cosmic rays are also present at other sites in the Galaxy, then low energy Galactic cosmic rays may play a very important role in the nucleosynthesis of the light isotopes $^6$Li, $^9$Be, $^{10}$B and $^{11}$B (Cassé, Lehoucq, & Vangioni-Flam 1995; Ramaty, Kozlovsky, & Lingenfelter 1996b).

In the present paper we briefly discuss these topics, referring the reader for more details to the papers mentioned above.

## 2. Solar Gamma Ray Spectroscopy

The solar flare gamma ray data is now sufficiently detailed to allow the conduct of a meaningful gamma ray spectroscopic analysis of the ambient solar atmosphere. The key is provided by the narrow line emission produced by accelerated protons and $\alpha$ particles interacting with ambient C and heavier nuclei. Owing to their narrower widths, these lines can be distinguished from the broader lines produced by accelerated C and heavier nuclei interacting with ambient H and He. The intensities of the narrow lines depend on the heavy element abundances and thus can be used to determine these abundances. Strong narrow line emission at 4.44, 6.13, 1.63, 1.37, 1.78, and 0.85 MeV, resulting from deexcitations in $^{12}$C, $^{16}$O, $^{20}$Ne, $^{24}$Mg, $^{28}$Si and $^{56}$Fe, respectively, has been observed from many flares. The most recent results, observed with the Solar Maximum Mission (SMM) from 19 flares (Share & Murphy 1995), allow the determination of the abundance ratios C/O, Mg/O and Mg/Ne for all 19 flares, Si/O for 14 flares and Fe/O for 12 flares (Ramaty et al. 1995a; 1996a).

Unlike atomic spectroscopy, nuclear spectroscopy does not require the temperature and ionic state of the ambient gas, neither of which are always well known. On the other hand, abundance determinations by nuclear spectroscopy do require information on the spectrum of the accelerated particles. For the SMM flare analysis (Ramaty et al. 1995a; 1996a) the accelerated particle spectra were constrained by using the 1.63 MeV $^{20}$Ne-to-6.13 MeV $^{16}$O and the 2.22 MeV neutron capture-to-4.44 MeV $^{12}$C line fluence ratios, both of which are strong functions of the particle spectrum.

As in other solar atmospheric abundance studies (e.g. Meyer 1992), it is useful to distinguish two groups of elements depending on their first ionization potential (FIP): low FIP (<10 eV) elements (Mg, Si and Fe) and high FIP (>11 eV) elements (C, O and Ne). The enhancement of low FIP-to-high FIP element abundance ratios in the corona relative to the photosphere is well established from atomic spectroscopy and solar energetic particle observations (e.g. Meyer 1992). Analysis of the gamma ray data has led to the following conclusions (Ramaty et al. 1995a; 1996a):

(i) For the high FIP elements C and O, the derived abundance ratio (by number) is $0.35 \lesssim C/O \lesssim 0.44$. This range is more consistent with the C/O = 0.43±0.05 given by Anders & Grevesse (1989) than with the C/O=0.48±0.1 of Grevesse & Noels (1992). But taking into account the large uncertainty of the latter, there is no real discrepancy. Furthermore, a single value of C/O is consistent with the data for all 19 flares, implying that C/O could have the same value throughout the gamma ray production region. This is in fact not surprising given that C/O is essentially the same in the photosphere and corona.

(ii) For another pair of high FIP elements, O and Ne, the gamma ray data is in better agreement with Ne/O=0.25 than with the commonly adopted photospheric and coronal value of 0.15. Such a low Ne/O could only be accommodated by a very steep accelerated particle spectrum which would take advantage of the very low threshold for the excitation of the 1.63 MeV level of $^{20}$Ne. The implied particle spectra, however, are too steep to produce sufficient neutrons to account for observations of the 2.22 MeV neutron capture line. In addition, the energy contained in ions with such steep spectra would be inconsistent with the overall flare energetics. Some EUV and X-ray observations also support a Ne/O higher than 0.15 (Saba & Strong 1993; Schmelz 1993; Widing & Feldman 1995).

(iii) To avoid values of Ne/O larger than 0.3 the accelerated particle energy spectra should be at least as steep as an unbroken power law down to about 1 MeV/nucl. For such power laws, the energy contained in the ions for the 19 analyzed flares ranges from about $10^{30}$ to well over $10^{32}$ ergs, and is thus comparable or even exceeds the energy contained in the nonrelativistic electrons that produce the hard X-rays in solar flares. Prior to these recent gamma ray analyses, it was widely believed that a large fraction of the released flare energy is contained in nonrelativistic electrons.

(iv) Considering the abundance ratios between elements of different FIP groups, both Mg/O and Mg/Ne show evidence for variability from flare to flare at about the $3\sigma$ level. For Mg/O this variation is confined to a range around the coronal value of 0.2 and does not go down to the photospheric value of 0.045. For Si/O and Fe/O the variations are also confined to a range around their respective coronal values. The fact that the low FIP-to-high FIP abundance ratios derived from gamma ray spectroscopy are enhanced relative to their respective photospheric values shows that the gamma ray production region lies above the photosphere.

## 3. On the Origin of the Light Isotopes

The discovery of gamma ray line emission from Orion (Bloemen et al. 1994) and the implied existence of large fluxes of low energy Galactic cosmic rays, has led to renewed discussions on the origin of the light elements. It has been known for over two decades that the relativistic Galactic cosmic rays (GCR) may have produced the observed solar system abundances of $^6$Li, $^9$Be and $^{10}$B (Meneguzzi, Audouze & Reeves 1971; Mitler 1972). These cosmic rays, however, cannot account for the abundances of $^7$Li and $^{11}$B. It is believed that most of the Galactic $^7$Li is produced in stars (e.g. Reeves 1994). Recent measurements of the boron isotopic ratio in meteorites yielded $^{11}$B/$^{10}$B values in the range 3.84 – 4.25 (Chaussidon & Robert 1995) which exceed the calculated GCR value by a factor of about 1.5. The implications of the Orion gamma ray observations on the origin of the light isotopes have been considered by Cassé, Lehoucq & Vangioni-Flam (1995) and Ramaty et al. (1996b). The advantages of producing the light isotopes with 'Orion-like' low energy cosmic rays are the following:

(i) Low energy cosmic rays can produce B such that $^{11}$B/$^{10}$B $\gtrsim$ 4. Energetic arguments favor light isotope production by low energy cosmic rays at typical particle energies around 20 MeV/nucl rather than at lower energies (Ramaty et al. 1996b). At these energies the excess $^{11}$B results mostly from $^{12}$C via the reactions $^{12}$C(p,pn)$^{11}$C and $^{12}$C(p,2p)$^{11}$B which have lower thresholds than the reaction $^{12}$C(p,2pn)$^{10}$B. At higher energies, and in reactions with $^{16}$O, the B isotopic ratio is significantly lower.

(ii) The low energy cosmic rays must be depleted in protons and $\alpha$ particles. The $\alpha$ particle depletion is necessary in order not to overproduce $^6$Li; the proton depletion ensures a linear dependence of the Be and B abundances on the Fe abundance in stars of various ages. If the low energy cosmic rays are poor in protons and $\alpha$ particles they will produce Be and B only from the breakup of accelerated C and O in interactions with ambient H and He; in this case both the target and projectile abundances could remain constant, leading to a linear growth of the Be and B abundances. On the other hand, the GCR would produce much of the isotopes from the breakup of C and O in the ambient medium whose abundances increase with time, leading to a quadratic growth.

(iii) Arguments of energetics have independently led to the suggestion that the low energy cosmic rays in Orion consist mostly of C and heavier nuclei with the protons and $\alpha$ particles strongly suppressed (Ramaty et al. 1995b; 1996b). This suppression could be the consequence of the particle injection process prior to the acceleration itself. The proposed injection sources are the winds of Wolf Rayet stars (Ramaty et al. 1995b), the ejecta of supernovae from massive star progenitors (Cassé et al. 1995; Ramaty et al. 1996b), and the pick up ions resulting from the breakup of interstellar grains (Ramaty et al. 1996b).